# Electronic Prescription System for Pediatricians

*Adebayo Omotosho, (PhD)*
Department of Computer Science
Landmark University, Omu Aran, Kwara State, Nigeria
*Ukeme Asanga, (B.Tech)*
*Aderogba Fakorede, (B.Tech)*
Department of Computer Science and Information Technology.
Bells University of Technology, Ota, Ogun State, Nigeria

**Abstract**

This paper presents the development of an electronic prescription system for pediatricians that considered the factors that influence a child's prescription. The system implements a knowledge base which contains drug information and formulary. It allows the pediatrician to have access to the electronic health record of patients before prescription writing. The resulting prescription is marked with verifiable randomly generated prescription ID before it is sent to the dispensing pharmacy and this accounts for the security feature of the prescription system. Microsoft Office Visio 2007, PHP and My SQL database server were used to present and develop the system. Implementation results showed the system is capable of reducing common prescription error as the most informed prescription is being generated for the child electronically.

**Keywords:** E-Prescription, pediatrics, pediatric specific dosage calculation, decision support

**Introduction**

Pediatrics is the only discipline that is concerned with the overall wellbeing (health, physical, mental, psychological growth and development) of infants, children, adolescents and offers them the opportunity to achieve full potential as an adult. The pediatric population includes children from birth to 18 years of age (WHO, 2007) and can be sub divided into neonates (birth to 1 month), infants (1 month to 2 years), children (2 years to 12 years) and adolescents (12 years to 18 years). They are more prone to the errors when using the paper prescription because, asides illegibility, certain calculations (based on the child's weight, height, age, body surface area, etc) need to be carried out to obtain dosage to be prescribed for a child and the importance of these calculations are often underemphasized and most times, the prescriber simply assumes a dosage for the child which could lead to cases of adverse drug effects since the child's body system is still developing unlike in adults where their system is fully developed (Kaushal et al, 2001; Alhanout et al, 2017).

Prescribing error rates in children were estimated to be between 5% and 27% in a systematic review by Miller et al (2007). A prescribing error can be defined as an incomplete or incorrect medication order that may result in adverse clinical conditions if administered as prescribed (Barber, Rawlins, & Franklin, 2003). In pediatrics prescribing, error rates vary and a study estimates prescribing error rates to be 13% of medication orders (Ghaleb et al, 2006). Prescribing errors are most prevalent with antibiotic agents but may also occur with medications that don't require weight-based dosing (Kaushal et al, 2010; Cresswell et al, 2016). Medication errors in children may lead to more severe complications due to the inability of some children to communicate adverse effects (Leyva, Sharif, & Ozuah, 2005). In addition, the paper/handwritten medical prescription process features some limitations like the vast quantity of types of medication, filing and handling of large volumes of documentation and low quality of handwritten annotation that serve as an incentive for computerization of the process called electronic prescription (e-prescription) which can manage prescription process electronically and sends it to the pharmacist for dispensing (Bell et al, 2004; Salmivalli & Hilmola, 2006).

E-prescription systems should allow pediatricians to be able to access knowledge base, patient's Electronic Health Record (EHR), prescription database to acquire knowledge concerning drugs, allergies, patient's past medical records, and also enables communication with the dispensing pharmacy to know if drug to be prescribed is available. This will ensure a more accurate and informed prescription for the patient hence, patient's overall health is improved. Since e-prescription is done electronically, it helps eliminate geographical distance between the child and the pediatrician hence providing the child with easier access to health care services. Many existing e-prescribing systems are not well designed for use in pediatric patients and lack the required features outlined (Johnson, Lehmann, & Council on Clinical Information Technology, 2013). Pediatric prescribing errors occur frequently and are not completely prevented by electronic prescribing systems. However, improvements such as minimized free-text entry, certain obligatory fields and integrated dose checking and indications in electronic prescription system for pediatric patients can improve the quality and efficiency of electronic prescribing in pediatrics (Maat et al, 2012).

Experts agree that medication error tends to cause more harm in the pediatric setting than within the adult population due to the need for weight-based dosing calculations, fractional dosing and the need for decimal points in the pediatric care setting (Landrigan, 2007). It is therefore necessary for health care professionals to pay attention to the challenges in the pediatric population. Research shows that the potential for adverse drug events within the in-patient pediatric population is three times more than that of hospitalized adults (Kaushal et al, 2001). The first study to develop and evaluate a trigger tool to detect Adverse Drug Effect (ADE) in an in-patient pediatric population identified an 11.1% of ADEs in pediatric patients, the study also showed that 22% of the ADEs were preventable, 17.8% could have been discovered earlier and 16.8% could have been handled more effectively (Takata et al, 2008). Children are more prone to prescribing errors and the resulting harm because most drugs used in childcare are primarily formulated for adults, most healthcare settings are built around the need of adults that is no well-trained pediatrician to attend to the needs of children (Kohn, Corrigan, & Donaldson, 2000), children are usually less able to physiologically tolerate medication errors due to still developing body systems, children may not be able to effectively communicate ADEs resulting from the use of a particular medication (Landrigan, 2007).

In 2006 and 2007, MEDMARX's database shows that close to 2.5% of pediatric medication errors led to patient harm with the most common types of harmful pediatric medication errors being improper dose/quantity (37.5%), omission error (19.9%), unauthorized/wrong drug (13.7%), and prescribing error (9.4%), also wrong administration technique, wrong time, drug prepared incorrectly, wrong dosage form, and wrong route. Medication errors involving pediatric patients were most often caused by performance deficit (43%), knowledge deficit (29.9%), procedure not followed (20.7%), miscommunication (16.8%), and calculation error, computer entry error, inadequate or lack of monitoring, improper use of pumps, and documentation errors. The MEDMARX Data Report reveals that approximately 32.4% of pediatric errors in the operating room involve an improper dose/quantity compared with 14.6 percent in the adult population and 15.4 percent in the geriatric population (Hicks et al, 2006).

E-prescription results in a quick and reliable prescription and also improves the prescribing process. A 2011 statistics show that 54% of physicians had already adopted the electronic health record (EHR) system and about three-quarters of these physicians reported that using their EHR system resulted in enhanced patient care (Dunlop, 2006). An inpatient CPOE system at Brigham and Women's Hospital in Boston following its implementation dropped the rates of serious medication errors (preventable ADEs + potential ADEs) by 55% (Bell et al, 2004). E-prescription reduces occurrence of errors, provides patients with improved health and quality of life, in the health care system, it reduces the shortage of professionals and also it has the potential of contributing to economic growth.

From the brief review, it is apparent that the pediatric population is vulnerable and at a greater risk of experiencing medication prescription errors due to the still developing body system of neonates, infants, children and adolescents/young adults. Electronic prescription system has not been fully integrated into the health care system and is not accepted by some physicians for certain reasons. It can be concluded that the use of an electronic system is still a rapidly developing field and even though it doesn't solve all the errors associated with medication prescriptions, it helps in improving the patients' health and safety, it's quick, convenient, reliable and overall helps in improving medication prescribing process (Ahmed et al. 2016).

**Materials and method**

The electronic prescription system is a client-server system with the front end or client-end of the system being used by the end users of the system to interact with the system and the back end or server end is used by the system to access necessary information.

**The e-prescription system and architecture**

The following subsection described individual modules of the system architecture. In addition, the randomly generated prescription ID in this system is a feature that helps ensure security of the prescription on the prescriber and pharmacist's end and reduces chances of forgery, to pick up medication so a person can't pick up someone else's medication. There is an option for printing in cases of network failure. The architecture in Figure 1 shows the system components and interactions.

**The application module**

This contains the pediatrician's workstation and the application software. The "Pediatrician's work station" is where the pediatrician works from to access the patient information, carry out electronic prescription and print it out for the patient or forward it to a registered pharmacy for dispensing. The "Application software" is installed on the pediatrician's workstation to enable him perform his duties properly, it is what gives the pediatricians' work station an edge over other systems being used by any random person. The application software authenticates and verifies the logon credentials of users that sign in to the system, allows you to add users and patients to the system, prescribe for the patients, allows pharmacist to check prescription to be dispensed to patient, etc.

**The knowledge base**

The "Knowledge base (drug information)" can also be referred to as the system's memory where patients' data and drug information are stored. The knowledge base stores all the necessary information about different drugs for the treatment of different disease. The main source of these data is Monthly Prescribing Reference (MPR) database for drug monographs available at http://www.empr.com/.

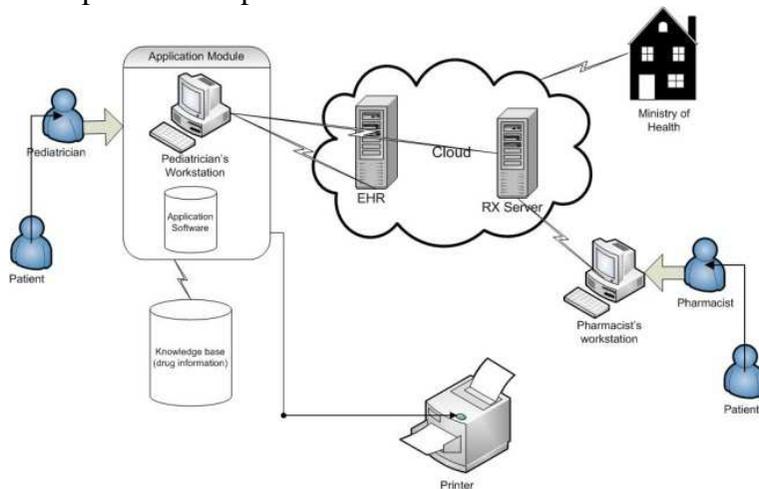

**Figure 1** Proposed system architecture for a secure electronic prescription system for pediatricians.

**The EHR**

It contains a database of the patient's health record. The EHR contains information like child's health record, family health record, allergy information, drug immunology record, cases of adverse drug reactions, etc. It serves as an electronic reference handbook.

**The prescription server**

This is a database which contains information about previous visits to the hospital, diagnosis, prescribed drugs and the effectiveness of the drugs in achieving the purpose for which it was prescribed. Prescriptions made by the pediatrician are being sent here so the dispensing pharmacy can access the prescription documents.

**The pharmacist's work station**

It is similar to the pediatrician's work station. The pharmacist's work station should be able to access some of the resources being used by the pediatrician's work station in prescribing. Its major work is to access the prescriptions made by the pediatrician and dispense the drug to the patient.

**Ministry of Health**

This allows for the ministry of health to access patient information stored in the electronic health record (EHR) of the patient and also in the prescription database to enable them carry out health related research, findings and recommendations on the general health. They can deduce the outbreak of a disease and come up with ways of rectifying and minimizing the occurrence of such diseases. There is also the printer that provides the option for pediatricians to print out prescription for patient to hand-deliver to dispensing pharmacy in the case of low network access or transmission error.

**System Modelling**

This was done using the Unified Modeling Language (UML) analysis model from Microsoft Office Vision 2007. The use case diagram and the activity diagram for the proposed architecture are represented in Figure 2 and Figure 3 respectively. Figure 2 shows the administrator, ministry of health, pharmacist, and pediatrician as the major actors of the system and also shows the roles of each of these actors within the system. The patient is also an actor within the system only he/she does not have direct access to the system. Figure 3 represents the activity diagram for the pediatrician and pharmacist and their functions.

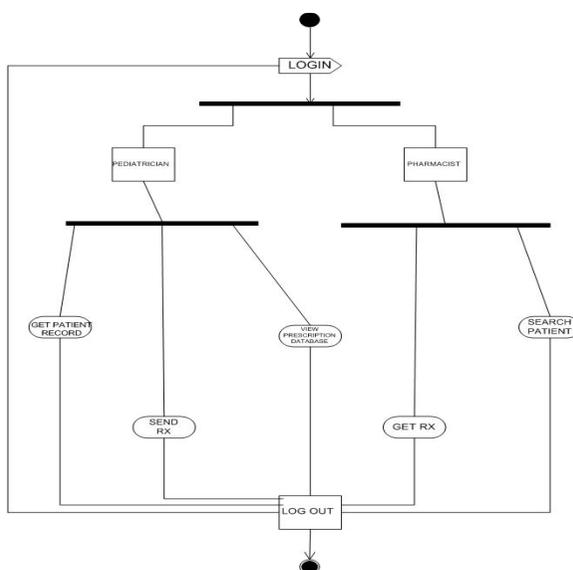

**Figure 2** Use case diagram of the e-prescription system.

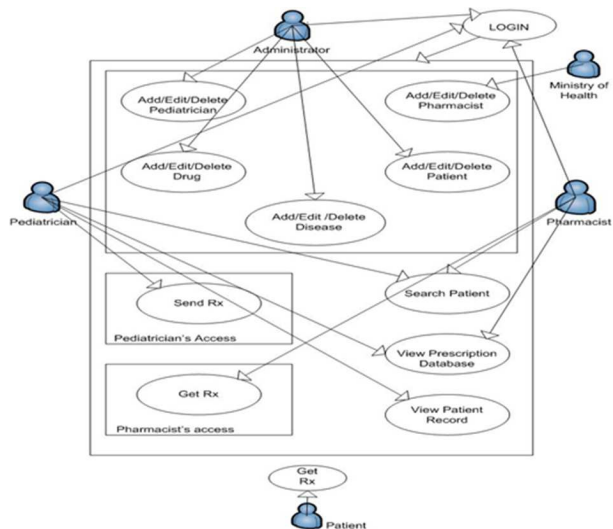

**Figure 3** Activity diagram of the e-prescription system

### System Implementation and Results

PHP programming language and MySQL database server were used for the development of the electronic prescription system for pediatricians. The figures representing the various functionalities of the electronic prescription system are thus described:

Figure 4 shows the logon authentication screen that permits only authorized pediatricians and pharmacists to login to the system and make use of the information on it. Here, the pediatrician or pharmacist inputs his or her username and password and clicks on login, the system then verifies the information to allow access to the system. In the case where wrong credentials are provided, the system returns an error message.

A patient that visits the clinic for consultation has to be registered on the system, the doctor then selects from the list of patients whom he want to prescribe for as shown in Figure 5. Where the patient to be prescribed for is not an in-patient or not found in the system database, a new patient registration page is launched so that the doctor can prescribe for him or her electronically.

Once the patient detail has been confirmed by the physician, a prescription can then be made by clicking on the "prescribe drugs" tab. The doctor's unique identifier is displayed throughout a particular prescription process. In Figure 6, Figure 7 and Figure 8 are the prescription pad the doctor writes his prescription on, from prescribe notes to drug selection and to sending of prescription. As soon as the doctor writes out the prescription on available spaces, he clicks done and the prescription is being sent to the prescription database and can also be printed out. The system also suggests drugs to be prescribed based on the diagnosis of the pediatrician and the system also computes the infant dosage after drug to be prescribed has been chosen by the prescriber using Clark's rule which is dependent on the child's body weight: Child's dose = (weight of child in pounds * adult dose)/150lbs (McMullan, Jones, & Lea, 2010). The prescriber can also decide to add more drugs by clicking on the add prescription tab. After filling out the prescription completely, the pediatrician sends it to a registered dispensing pharmacy or prints it out for patient to hand deliver.

In Figure 9, there is a display of all possible drug information needed for prescription from MPR database. The prescriber can make reference to the drug information of any drug he or she wants to prescribe and make the best choice of the possible drugs that can be used.

Figure 10 displays the pharmacist login to the system where the random prescription ID generated for the patient has to be entered into the system to be able to access the document and confirm that a particular prescription belongs to a particular patient so that one cannot pick up some else's prescription and also the validity of the prescription can be checked. The validity period here being three days after the drug was prescribed.

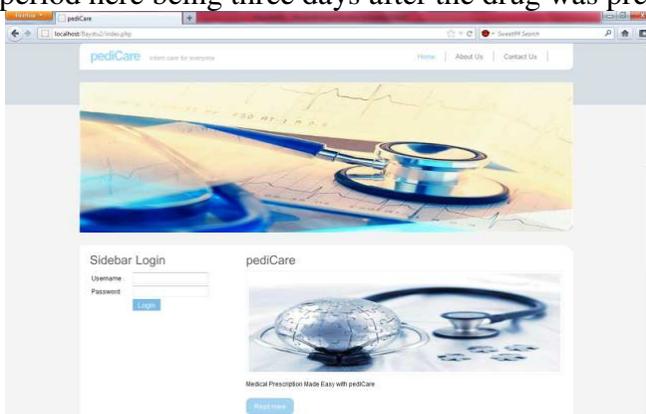

**Figure 4.** Logon authentication page

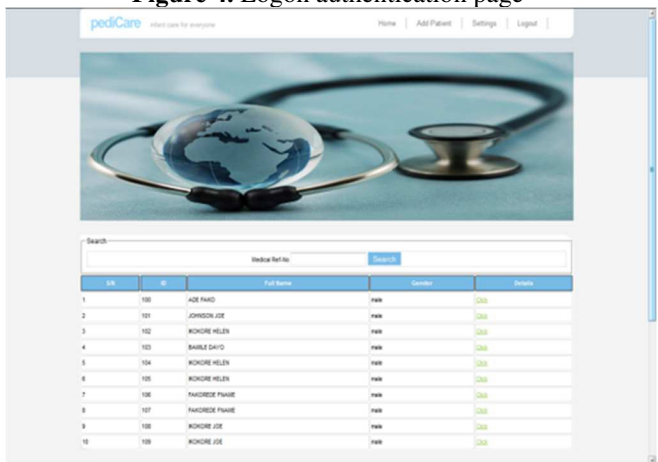

**Figure 5.** Doctor selecting patient

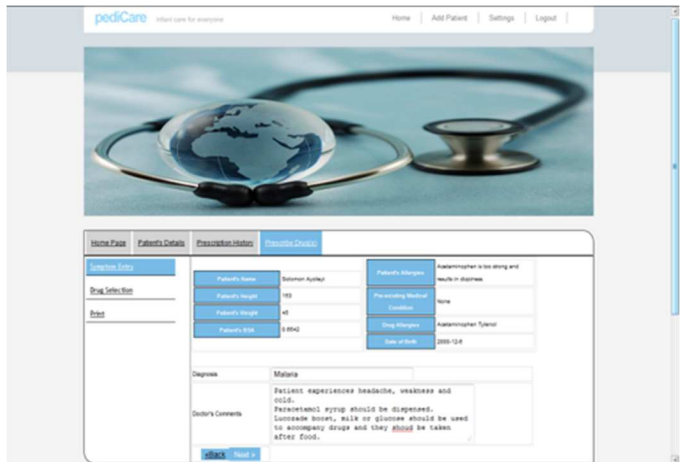

**Figure 6**. Symptoms entry

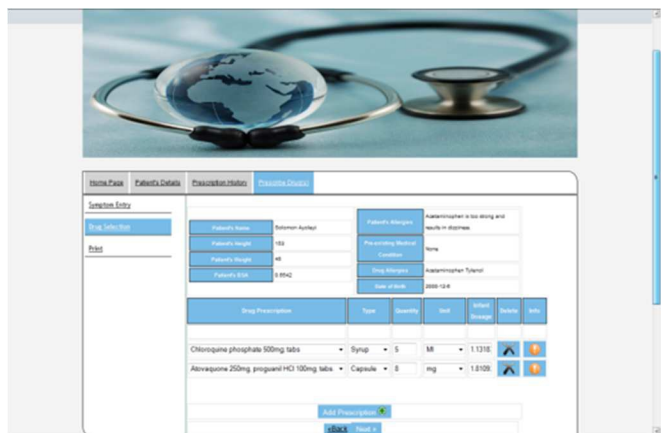

**Figure 7**. Drug Selection

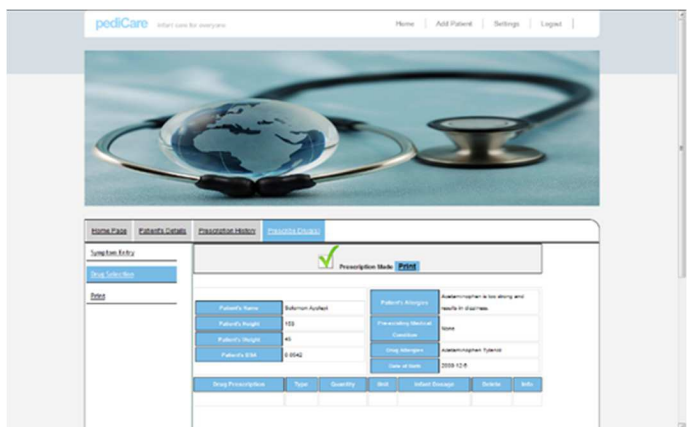

**Figure 8**. Prescription page of the doctor

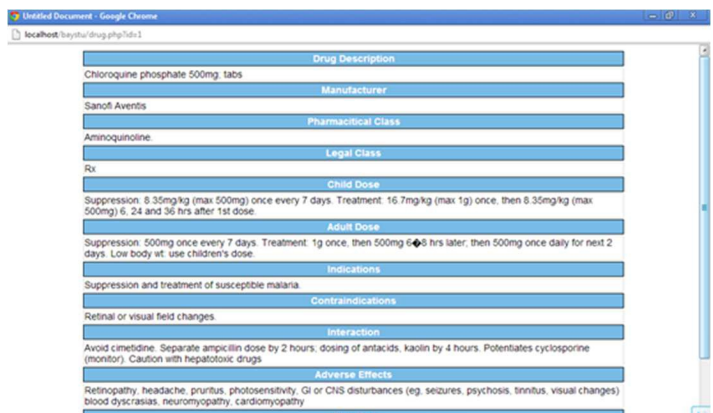
**Figure 9**. Drug information

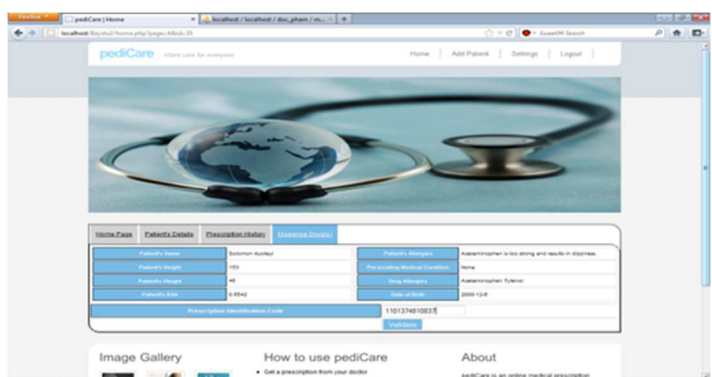
**Figure 10**. Pharmacist validating prescription

**Conclusion**

Research has proven that the employment of the electronic prescription system in the health sector does not only reduce errors of handwriting but also provides a more accurate prescription for the patient. The e-prescription system presented in this study is capable of providing pediatrician with features that makes use of patient's records and other knowledge available to generate a well-informed prescription. The prescriber's identification number on the prescription document enhances the security feature of the system. Future work will consider carrying out a performance evaluation on this electronic prescription system and make the system compatible with mobile platforms.

**Acknowledgments**